\documentclass[fleqn,twoside]{article}
\usepackage{espcrc2}
\usepackage{amssymb}
\usepackage{amsbsy}
\usepackage{amsfonts}

\usepackage{graphicx}
\newcommand{\AmS}{{\protect\the\textfont2
A\kern-.1667em\lower.5ex\hbox{M}\kern-.125emS}}

\newcommand{\be}{\begin{equation}}
\newcommand{\ee}{\end{equation}}
\newcommand{\bea}{\begin{eqnarray}}
\newcommand{\eea}{\end{eqnarray}}

\newcommand{\eqref}[1]{(\ref{#1})}

\title{
\vspace{-5mm}
\rightline{\small MPP-2004-107}
Finite size mass shift formula for stable particles revisited}

\author{
Yoshiaki Koma\thanks{Talk presented by Y. Koma.%
}and Miho Koma\\[0.2cm]
Max-Planck-Institut f\"ur Physik,
F\"ohringer Ring 6, D-80805, M\"unchen, Germany
}

\begin{document}

\begin{abstract}
L\"uscher's finite size mass shift formula 
in a periodic finite volume, involving
forward scattering amplitudes
in the infinite volume, is revisited
for the two stable distinguishable particle system.
The generalized mass shift formulae 
for the boson and fermion are
derived in the boson-boson and fermion-boson
systems, respectively.
The nucleon mass shift is discussed
in the nucleon-pion system.
\end{abstract}

\maketitle

\section{Introduction}
\label{sec:intro}

\par
Nowadays the control of finite volume effects 
in lattice QCD simulations with dynamical fermions
becomes a more important issue in order to 
determine the hadron spectrum precisely.
Applications of chiral perturbation theory
(ChPT)~\cite{Becher:1999he} to the measured
spectrum on the lattice have been done 
not only to achieve the chiral 
extrapolation~\cite{Bernard:2003rp,Procura:2003ig},
but also to find its finite volume dependence
towards the thermodynamic 
limit~\cite{Orth:2003nb,Colangelo:2003hf,%
AliKhan:2003cu,Beane:2004tw}.

\par
In this context, L\"uscher's formula, relating the 
mass shift in finite volume with periodic boundary 
conditions to forward elastic scattering amplitudes 
in infinite volume,
provides us with an elegant tool for such a 
purpose~\cite{Luscher:1983rk}.
L\"uscher presented
a rigorous proof of his formula for the case of a 
self-interacting boson to all orders in perturbation 
theory~\cite{Luscher:1986dn}.

\par
Among the applications of ChPT,
the QCDSF-UKQCD collaboration estimated the 
finite volume effect on the nucleon mass from data
in $N_{f}=2$ lattice QCD~\cite{AliKhan:2003cu},
applying the mass shift formula derived within ChPT.
Along the way, they found however, 
when expressing their formula in terms of 
$F_{N \pi}(\nu)$ in the same order of ChPT,
that the factor of the pole term 
(the first term of Eq.~\eqref{eqn:nucleon-massshift} below) 
is twice larger than that of L\"uscher's 
in Ref.~\cite{Luscher:1983rk}.
There seems to be no mistake in the 
formula in Ref.~\cite{AliKhan:2003cu}, at least
within the infrared regularization scheme~\cite{Becher:1999he}.
On the other hand, L\"uscher's formula being considered as general
such that it can also be applicable to ChPT at any orders,
this discrepancy poses a structural question.

\par
In Ref.~\cite{Koma}, we thus investigated 
the mass shift formula 
for the interacting two stable particle system
along the lines of L\"uscher's proof for a 
self-interacting bosonic theory.
In this report, we present the
resulting formulae
and an application to the nucleon mass shift in the
$N$-$\pi$ system.

%%%%%%%%%%%%%%%%%%%%%%%%%%%%%%%%%%
\section{Finite size mass shift formulae}
\label{sec:result}

\par
The physical mass of a stable particle is given by the position
of the pole of the propagator of an asymptotic field.
In the framework of  perturbation theory the pole is
shifted from the bare one due to the self energy
arising from virtual polarization effects.
In finite volume, the expressions for the self energy involve sums
over discrete spatial loop momenta, 
$\vec{q}(L) = 2\pi \vec{n}/L$ $(\vec{n} \in Z\!\!\!Z^{3})$.
By using the Poisson summation formula, 
such a summation can be rewritten  as
an integral with another summation over integer vectors 
$\vec{m} \in Z\!\!\!Z^{3}$
and an exponential factor:
\bea
&& \frac{1}{L^{3}} \sum_{\vec{n} \in Z\!\!\!Z^{3}}
\int \frac{dq_{0}}{2 \pi}  f (q_{0},\vec{q}(L) )\nonumber\\*
&&
=
\sum_{\vec{m} \in Z\!\!\!Z^{3}}
\int \frac{d^{4}q}{(2 \pi)^{4}} e^{-iL \vec{m}\cdot \vec{q}}
f (q_{0},\vec{q} \; )\; ,
\label{eqn:poisson}
\eea
where $f (q)$ is a function composed of
 propagators and vertex functions.
Then, the difference of the self energies
between the finite and infinite volumes,
appearing in the definition of the mass shift, 
can be defined by the sum over $|\vec{m}| \ne 0$,
since 
$|\vec{m}|=0$ corresponds to the integral in infinite volume.
The asymptotic formula at large $L$ is
given by the contribution of~$|\vec{m}| =1$.

\par
Let $M(L)$ and $m$ be the masses in finite and infinite volumes,
respectively.
The boson mass shift formula for $\phi_{A}$ in the 
$\phi_{A}$-$\phi_{B}$ system,
$\Delta m_{A} (L) =M_{A} (L) - m_{A}$,
is found to be\footnote{
we assume that only $A$ particle carries a conserved
charge.}
\bea
&&\!\!\!\!\!\!\!\!\!
\Delta m_{A} (L)
=
 -   \frac{3}{8\pi m_{A} L }
\Biggl [
\frac{ \lambda_{\mathit{AAB}}^{2}}{ 2 \nu_{B}}
e^{- L \sqrt{m_{B}^{2}-\nu_{B}^{2}}}
\nonumber\\*
&&\!\!\!\!\!\!\!\!\!
+ \int_{-\infty}^{\infty}\!\!  \frac{dq_{0}}{2 \pi}
% \; 
e^{-L\sqrt{m_{A}^{2}+q_{0}^{2}}} F_{\mathit{AA}}(i q_{0})
\nonumber\\*
&&\!\!\!\!\!\!\!\!\!
+ \int_{-\infty}^{\infty} \!\! \frac{dq_{0}}{2 \pi}
% \; 
e^{-L \sqrt{m_{B}^{2}+q_{0}^{2}}} F_{\mathit{AB}}(iq_{0})
\Biggr ] 
\!\!+ \! O( e^{- L\bar{m}}) \; ,
\label{eqn:massshift-boson}
\eea
where $F_{\mathit{AB}}(\nu)$ and $F_{\mathit{AA}}(\nu)$ 
denote the forward scattering amplitudes of the processes
$A+B \to A+B$ and $A+A \to A+A$ 
in the infinite volume, respectively.
$\lambda_{\mathit{AAB}}$ 
is an effective renormalized coupling defined from the residue 
of $F_{\mathit{AB}}(\nu)$ at $\nu= \pm \nu_{B} 
= \pm m_{B}^{2}/2m_{A}$ as
\bea
\lim_{\nu \to \pm \nu_{B}}(\nu^{2}-\nu_{B}^{2}) \; 
F_{\mathit{AB}}(\nu)
=\frac{\lambda_{\mathit{AAB}}^{2}}{2}\; .
\label{eqn:eff-re-coupling}
\eea
The error term is defined by $\bar{m} \ge 
\sqrt{2(m_{B}^{2}-\nu_{B}^{2})}$, which is 
due to the neglect of~$|\vec{m}| \geq 2$ contributions.
The formula is valid for $m_{A}\geq m_{B}$.
However, if $m_{A}\gg m_{B}$, we may neglect the second term,
since the contribution becomes smaller 
than the error term.

\par
The fermion mass shift formula for $\Psi_{A}$
in the $\Psi_{A}$-$\phi_{B}$ system 
($m_{A}\geq m_{B}$) is found to be
\bea
&&\!\!\!\!\!\!\!\!\!
\Delta m_{A} (L)
=
-
\frac{3}{8\pi m_{A} L}
\Biggl [
\frac{\lambda_{\mathit{AAB}}^{2}}{2 \nu_{B}}
e^{- L\sqrt{m_{B}^{2}-\nu_{B}^{2}}}
\nonumber\\*
&&\!\!\!\!\!\!\!\!\!
- \int_{-\infty}^{\infty} \!\frac{dq_{0}}{2 \pi}
% \; 
e^{-L\sqrt{m_{A}^{2}+q_{0}^{2}} }
\{ F_{\mathit{AA}}(iq_{0})+F_{\mathit{A\bar{A}}}(iq_{0}) \}
\nonumber\\*
&&\!\!\!\!\!\!\!\!\!
+
\int_{-\infty}^{\infty} \! \frac{dq_{0}}{2 \pi}
% \; 
e^{-L\sqrt{m_{B}^{2}+q_{0}^{2}}} F_{\mathit{AB}}(iq_{0})
\Biggr ] 
\!\!+ \! O(e^{-L\bar{m}}) \; ,
\label{eqn:massshift-fermion}
\eea
where three types of the forward scattering amplitudes 
contribute to the formula,
$F_{\mathit{AB}}(\nu)$: $A+B \to  A+B$,
$F_{\mathit{AA}}(\nu)$: $A+A \to A+A$ 
and
$F_{\mathit{A\bar{A}}}(\nu)$: 
$A+\bar{A}\to A+\bar{A}$.
Apart from the relative minus sign in front of the second
term, which is due to Fermi statistics, the 
formula is almost the same as 
Eq.~\eqref{eqn:massshift-boson}.

%%%%%%%%%%%%%%%%%%%%%%%%%%%%%%%
\section{The nucleon mass shift}
\label{sec:nucleon}

As an application of 
Eq.~\eqref{eqn:massshift-fermion},
let us discuss the nucleon mass shift in the $N$-$\pi$ system.
Since the formula is expected to hold 
nonperturbatively,
it is interesting to estimate the 
mass shift by inserting the $N$-$\pi$ scattering
amplitude which is known from experiment.
Therefore, the following analysis
can be regarded as an estimate of the finite volume
effect on the nucleon mass 
when the realistic pion mass is achieved in lattice
QCD simulations with dynamical fermions.

\par
According to H\"ohler~\cite{Hohler:1984ux}, the 
subthreshold expansion of the 
$N$-$\pi$ forward scattering amplitude 
around $\nu = 0$ is parametrized as
\bea
&&D^{+} (\nu)
= 
\frac{g^{2}}{m_{N}}\frac{\nu_{B}^{2}}{\nu_{B}^{2}-\nu^{2}}
+d_{00}^{+} \; m_{\pi}^{-1} 
\nonumber\\*
&&
+d_{10}^{+} \; m_{\pi}^{-3} \nu^{2}
+d_{20}^{+} \; m_{\pi}^{-5} \nu^{4} + O(\nu^{6}) \; .
\label{eqn:d-plus}
\eea
Here $m_{N}=938$ MeV and $m_{\pi}=140$ MeV are the masses 
of the nucleon and pion ($m_{\pi}/m_{N}=0.149$), 
and $g^{2}/4 \pi =14.3$.
The effect of isospin symmetry breaking is neglected.
The first term is identified with the pseudovector nucleon Born term
with $\nu_{B}=m_{\pi}^{2}/2m_{N} \approx 0.07 m_{\pi}$.
The coefficients of the other terms are given by 
$d_{00}^{+}=-1.46(10)$, $d_{10}^{+}=1.12(2)$ and 
$d_{20}^{+}=0.200(5)$~\cite{Hohler:1984ux}.
We only take into account the mean of these values.

\par
By sandwiching $D^{+}(\nu)$ between the nucleon spinors 
$\bar{u}$ and $u$
and taking into account the isospin factor, we can relate this to  
the amplitude $F_{AB}(\nu) \equiv F_{N\pi}(\nu)$ in 
Eq.~\eqref{eqn:massshift-fermion} by
$F_{N\pi}(\nu) = 6 m_{N} D^{+}(\nu)$.
The effective coupling is then computed by
using Eq.~\eqref{eqn:eff-re-coupling} as
$\lambda_{\mathit{NN\pi}}^{2} = -12 g^{2} \nu_{B}^{2}$.
In this case, since $m_{N} \gg m_{\pi} $, the second term in
Eq.~\eqref{eqn:massshift-fermion} can be neglected.
The mass shift formula, divided by 
the nucleon mass itself, is reduced to
\bea
&&\!\!\!\!\!\!\!\!\!
\delta (\xi \! = \! Lm_{\pi}) 
\equiv  \Delta m_{N}/m_{N} 
\nonumber\\*
&&\!\!\!\!\!\!\!\!\!\approx
\frac{9}{2 \xi} \left (\frac{g^{2}}{4 \pi} \right )
\left (\frac{m_{\pi}}{m_{N}} \right )^{3} 
e^{- \xi \sqrt{1- \nu_{B}^{2}/m_{\pi}^{2}}}
\nonumber\\*
&&\!\!\!\!\!\!\!\!\!
- \frac{3}{16 \pi^{2} \xi} 
\left (\frac{m_{\pi}}{m_{N}} \right )^{2} \!\!
\int_{-\infty}^{\infty}dy \;
e^{-\xi \sqrt{1+y^{2}}} F_{N\pi} (i m_{\pi} y) \nonumber\\*
&&\!\!\!\!\!\!\!\!\! =
\delta_{P}(\xi)+\delta_{B}(\xi)+\delta_{R} (\xi)\; ,
\label{eqn:nucleon-massshift}
\eea
where $\delta_{P}(\xi)$ is the pole term,
and $\delta_{B}(\xi)$ and $\delta_{R}(\xi)$ 
correspond to the contributions of the pseudovector 
Born term
and the rest in Eq.~\eqref{eqn:d-plus}.
This expression is consistent with the formula
given in Ref.~\cite{AliKhan:2003cu}
by the QCDSF-UKQCD collaboration with ChPT.
It means that the pole term has been 
underestimated by factor two in L\"uscher's formula in 
Ref.~\cite{Luscher:1983rk}.

\begin{figure}[!t]
\centering\includegraphics[width=7.7cm]{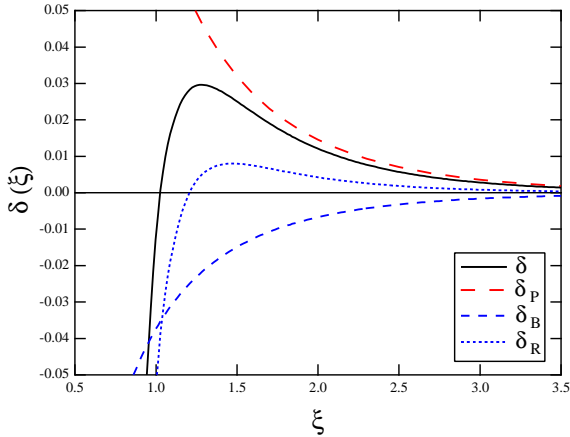}\vspace*{-0.5cm}
\caption{The nucleon mass shift as a function of $\xi=L m_{\pi}$.} 
\label{fig:delta_xi}
\vspace*{-0.5cm}
\end{figure}

\par
We plot $\delta (\xi)$ in Fig.~\ref{fig:delta_xi},
where $\xi =1$ corresponds to $L=1.4$ fm.
We find that $\delta (\xi)$ suffers strongly from higher 
order contributions of $\nu$ in the range $\xi \leq 1$.
For instance, $\delta_{R} (\xi)$ causes the negative mass 
shift within the leading mass shift formula ($|\vec{m}|=1$).
In this range, the contribution from $|\vec{m}| >1$ 
to the formula, of course, will not be negligible.
On the other hand, $\delta (\xi)$ seems to be 
mostly described by $\delta_{P}(\xi)$ as $\xi$ increases.
However this is due to the cancellation between
$\delta_{B}(\xi)$ and $\delta_{R}(\xi)$.

\section{Summary}
\label{sec:summary}

We have studied the general finite size mass shift formula 
for the two stable distinguishable particle system
in a periodic finite volume along the lines of L\"uscher's proof
for an identical bosonic theory.
The main results are Eqs.~\eqref{eqn:massshift-boson}
and~\eqref{eqn:massshift-fermion}.

\par
The overall error terms of the formulae can probably be reduced 
by taking into account the summation over integer 
vectors $\vec{m}$
in Eq.~\eqref{eqn:poisson} 
without modifying the derivation,
although a complete knowledge of
the analyticity properties of the vertex functions is needed
to control the final error term.

\par
There are now valid finite size mass shift formulae 
for the two particle systems, in addition to that for the identical 
bosonic system~\cite{Luscher:1986dn}.
For every case all these formulae are obtained in the same way
by carefully analyzing the appropriate set of 
self-energy diagrams.

\section*{Acknowledgments}

We are grateful to P. Weisz for introducing us to this interesting 
topic and also for numerous discussions during the course of the 
present work.
We also appreciate useful comments from M. L\"uscher
and G. Colangelo.
We are partially  supported  by the
DFG Forschergruppe `Lattice Hadron Phenomenology.'
M.K. is also supported by Alexander von Humboldt
foundation, Germany.

\end{document}